\documentstyle[preprint,aps]{revtex}
\newcommand{\bce}{\begin{center}}
\newcommand{\ece}{\end{center}}
\newcommand{\beq}{\begin{equation}}
\newcommand{\eeq}{\end{equation}}
\newcommand{\bea}{\vspace{0.25cm}\begin{eqnarray}}
\newcommand{\eea}{\end{eqnarray}}

\newcommand{\ba}{\begin{array}}
\newcommand{\ea}{\end{array}}

\newcommand{\r}{\mbox{{\boldmath $\rho$}}}
\newcommand{\ta}{\mbox{{\boldmath $\tau$}}}

\newcommand{\bfb}{\mbox{{\boldmath $\beta$}}}
\newcommand{\bfa}{\mbox{{\boldmath $\alpha$}}}

\def\lsim{\mathrel{\rlap{\lower4pt\hbox{\hskip1pt$\sim$}}
    \raise1pt\hbox{$<$}}}	  
\def\gsim{\mathrel{\rlap{\lower4pt\hbox{\hskip1pt$\sim$}}
    \raise1pt\hbox{$>$}}}	  

\def\beq{\begin{equation}}
\def\endeq{\end{equation}}
\def\bea{\begin{eqnarray}}
\def\arr{\begin{eqnarray}}
\def\eea{\end{eqnarray}}

\def\q2{$Q^{2}$}
\def\s2{2$S$}

\makeindex
\begin{document}
\draft
\title{
Fully quantum treatment of the Landau--Pomeranchuk--Migdal effect
in QED and QCD}

\author { B.G.Zakharov}
\address{
L.D.Landau Institute for Theoretical Physics,
GSP-1, 117940,\\
Ul. Kosygina 2, V-334 Moscow, Russian Federation
}

\date{\today}

\maketitle
\begin{abstract}
For the first time a rigorous quantum treatment of
the Landau-Pomeranchuk-Migdal effect in QED and QCD is given.
The rate of photon (gluon) radiation
by an electron (quark) in medium is expressed through
the Green's function of a two-dimensional Schr\"odinger
equation with an imaginary potential. In QED this potential
is proportional to the dipole cross section for scattering
of $e^{+}e^{-}$ pair off an atom, in QCD it is
proportional to the cross section of interaction
with color centre of the color singlet
quark-antiquark-gluon system.

\end{abstract}
\newpage


The effect of multiple scattering on bremsstrahlung
(the Landau-Pomeranchuk-Migdal (LPM) effect \cite{LP,Migdal})
in QED and QCD has recently
attracted much attention \cite{GW,BD}. However,
a rigorous treatment of the LPM effect for
an arbitrary energy of the photon (gluon) is
as yet lacking.
In the present paper, based on the technique of Ref. \cite{BGZ},
 we develop a quantum theory of the LPM
effect for the whole range of photon (gluon) energies.
As usual, we treat the medium
as a system of uncorrelated static scatterers (atoms).

In order to set the background we start with the LPM effect in QED.
In vacuum the evolution of the wave
function without radiative corrections of a relativistic electron
with longitudinal momentum $p_{z}$ in the variable
$\tau=(t+z)/2$ at $(t-z)=$const
is described by the
Hamiltonian $H=({\bf p}^{2}+m^{2}_{e})/2\mu_{e}$ \cite{Bjorken},
where $\bf{p}$ is a transverse
momentum, $m_{e}$ is the electron mass, and $\mu_{e}=p_{z}$.
Consequently,
the electron wave functions
at planes $z=z_{1}$ and
$z=z_{2}$ are related by
\beq
\psi(\r_{2},z_{2})=\int
d\r_{1}K_{e}(\r_{2},z_{2}|\r_{1},z_{1})
\psi(\r_{1},z_{1})\,,
\label{eq:1}
\eeq
where $\r_{1,2}$ are the transverse coordinates,
and $K_{e}$ is the
Green's function of the two-dimensional
Schr\"odinger equation for a particle with mass $\mu_{e}\,$,
times the phase factor $\exp[-i m^{2}_{e}(z_{2}-z_{1})/2\mu_{e}]$.
Eq.~(\ref{eq:1}) holds for each helicity state.
At high energy spin effects in interaction of an electron
with an atom vanish, and equation analogous
to (\ref{eq:1}) holds for propagation of an electron
through medium as well. The corresponding propagator reads
\beq
K_{e}(\r_{2},z_{2}|\r_{1},z_{1})=
\int {\cal D}\r
\exp
\left\{
i\int
dz
\left[
\frac{\mu_{e} \dot{\r}^{2}}{2}+e\, U(\r,z)
\right]-\frac{i m_{e}^{2}(z_{2}-z_{1})}{2\mu_{e}}
\right\}\,\,,
\label{eq:2}
\eeq
where $\dot{\r}=d\r/dz$, $U(\r,z)$ is the potential of the medium.

The interaction of an electron with the photon field
generates the radiative correction, $\delta K_{e}$, to
the propagator (\ref{eq:2}).
To leading order in $\alpha=1/137$,
$\delta K_{e}$ is generated
by sequential transitions $e\rightarrow e'\gamma\rightarrow e$.
The probability of passage of an electron through the target
without
radiation of the photon can be written in terms of $K_{e}$ and
$\delta K_{e}$ as
\beq
P_{e}=1+2\mbox{Re}\int d \r_{1}
 d \r_{1}'
d \r_{2}
\left[\langle
\delta K_{e}(\r_{2},z_{2}|\r_{1},z_{1})
K^{*}_{e}(\r_{2},z_{2}|\r_{1}',z_{1})
\rangle
-({\rm vac})\right]
\,,
\label{eq:3}
\eeq
where $\langle\,...\,\rangle$ means averaging over
the states of the target, and (vac) denotes the
vacuum $\delta K_{e} K^{*}_{e}$ term. The subtraction of the vacuum
term takes
into account the renormalization of the electron
wave function.
Eq.~(\ref{eq:3}) is written for
the target with the density
independent of $\r$, and the points
$z_{1}$ and $z_{2}$ are assumed to be at large
distances before and after the target, respectively. The initial
electron flux is normalized to unity.
Evaluation of $P_{e}$ allows to determine the probability
of radiation of the photon, $P_{\gamma}$,
through the unitarity relation
$P_{e}+P_{\gamma}=1$.

The contribution of transitions
$e\rightarrow e'\gamma\rightarrow e$ to
$\delta K_{e}$ can be written as
\begin{eqnarray}
\delta K_{e}(\r_{2},z_{2}|\r_{1},z_{1})=-
\int\limits_{0}^{1} dx
\int\limits_{z_{1}}^{z_{2}} d \xi_{1}\int\limits_{\xi_{1}}^{z_{2}}
d \xi_{2}
\int d \ta_{1}d \ta_{2}
g(\xi_{1},\xi_{2},x)\nonumber\\
\times
K_{e}(\r_{2},z_{2}|\ta_{2},\xi_{2})
K_{e'}(\ta_{2},\xi_{2}|\ta_{1},\xi_{1})
K_{\gamma}(\ta_{2},z_{2}'|\ta_{1},\xi_{1})
K_{e}(\ta_{1},\xi_{1}|\r_{1},z_{1})\,.
\label{eq:4}
\end{eqnarray}
Here
the indices $e'$ and $\gamma$ label the
electron and photon propagators
for the intermediate $e'\gamma$ state.
The transverse masses which enter $K_{e'}$ and $K_{\gamma}$ are
given by $\mu_{e'}=(1-x)\mu_{e}$ and $\mu_{\gamma}=x\mu_{e}$, where
$x$ is the light-cone Sudakov variable of the photon.
The vertex function $g(\xi_{1},\xi_{2},x)$ equals
\beq
g(\xi_{1},\xi_{2},x)=\Lambda_{nf}(x)
[{\bf v}_{\gamma}(\xi_{2})
-{\bf v}_{e'}(\xi_{2})]
\cdot
[{\bf v}_{\gamma}(\xi_{1})
-{\bf v}_{e'}(\xi_{1})]
+\Lambda_{sf}(x)\,,
\label{eq:5}
\eeq
where
$\Lambda_{nf}(x)={\alpha[4-4x+2x^{2}]}/{4x}$,
$\Lambda_{sf}(x)=
\alpha m_{e}^{2}x/2\mu_{e'}^{2}$,
${\bf v}_{\gamma}$ and ${\bf v}_{e'}$ are the
transverse velocity operators, which act on the
corresponding propagators in Eq.~(\ref{eq:4}). The two terms in (\ref{eq:5})
correspond to the $e\rightarrow e'\gamma$ transitions
conserving (nf) and changing (sf) the electron helicity.

Making use of Eqs.~(\ref{eq:2})-(\ref{eq:4}) we can write the rate of
the bremsstrahlung in the following differential form
(we suppress the vacuum term, which will be recovered in the final
formula (\ref{eq:16}))
\bea
\frac{d P_{\gamma}}{d x}=2\mbox{Re}
\int\limits_{z_{1}}^{z_{2}} d \xi_{1}
\int\limits_{\xi_{1}}^{z_{2}}d \xi_{2}\int
d \r_{1} d\r_{1}' d\ta_{1}d\ta_{1}'
d\ta_{2}d\ta_{2}'d\r_{2}
g(\xi_{1},\xi_{2},x)\nonumber\\
\times
S(\r_{2},\r_{2},z_{2}|\ta_{2},\ta_{2}',\xi_{2})
M(\ta_{2},\ta_{2}',\xi_{2}|\ta_{1},\ta_{1}',\xi_{1})
S(\ta_{1},\ta_{1}',\xi_{1}|\r_{1},\r_{1}',z_{1})\,,
\label{eq:6}
\end{eqnarray}
where $S$ and $M$ are defined as
\beq
S(\r_{2},\r_{2}',\xi_{2}|\r_{1},\r_{1}',\xi_{1})
=\int{\cal D}\r_{e}{\cal D}\r_{e}'
\exp\left[\frac{i\mu_{e}}{2}\int d\xi(\dot{\r}_{e}^{2}
-\dot{\r}_{e}'^{2})\right]
\Phi(\{\r_{e}\},\{\r_{e}'\})\,,
\label{eq:7}
\eeq
\bea
M(\r_{2},\r_{2}',\xi_{2}|\r_{1},\r_{1}',\xi_{1})=
\int {\cal D}\r_{e}{\cal D}\r_{e'}{\cal D}\r_{\gamma}
\exp\left\{
\frac{i}{2}\int d \xi
(\mu_{e'}\dot{\r}_{e}^{2}+\mu_{\gamma}\dot{\r}_{\gamma}^{2}
-\mu_{e}\dot{\r}_{e}^{2})\right.\nonumber\\
\left.
-\frac{i(\xi_{2}-\xi_{1})}{l_{f}}\right\}
\Phi(\{\r_{e'}\},\{\r_{e}\})\,,
\label{eq:8}
\eea
\beq
\Phi(\{\r_{i}\},\{\r_{j}\})=
\langle\exp\left\{ie\int d \xi
[U(\r_{i}(\xi),\xi)-U(\r_{j}(\xi),\xi)]\right\}\rangle\,.
\label{eq:9}
\eeq
Here
$
l_{f}=2\mu_{e}(1-x)/m_{e}^{2}x
$ is usually called the photon formation length.
The boundary conditions for trajectories in Eq.~(\ref{eq:7})
are $\r_{e}(\xi_{1,2})=\r_{1,2}$,
$\r_{e}'(\xi_{1,2})=\r_{1,2}'$,  and in Eq.~(\ref{eq:8})
$\r_{e',\gamma}(\xi_{1,2})=\r_{1,2}$,
$\r_{e}(\xi_{1,2})=\r_{1,2}'$.
Averaging over positions of atoms in Eq.~(\ref{eq:9})
yields \cite{BGZ}
\beq
\Phi(\{\r_{i}\},\{\r_{j}\})=
\exp\left[-\frac{1}{2}\int
d\xi n(\xi)\sigma(|\r_{i}-\r_{j}|)\right]\,,
\label{eq:10}
\eeq
here $n(\xi)$ is the target density, and
$\sigma(\rho)$ is the dipole cross section for
scattering of the $e^{+}e^{-}$ pair of transverse size $\rho$
on a free atom. For the atomic potential
$\phi(r)=4\pi(Z\alpha/r)\exp(-a/r)$
($a=1.4Z^{-1/3}r_{B}$) $\sigma(\rho)$ is given by
\beq
\sigma(\rho)=8\pi(Z\alpha a)^{2}
\left[1-\frac{\rho}{a}K_{1}\left(\frac{\rho}{a}\right)\right]\,.
\label{eq:11}
\eeq
For $\rho\ll a$,
which will be important in the considered problem,
$
\sigma(\rho)
\simeq C(\rho)\rho^{2}\,,
$
where
\beq
C(\rho)=4\pi(Z\alpha)^{2}
\left[\log\left(\frac{2a}{\rho}\right)+\frac{(1-2\gamma)}{2}\right]\,,
\;\;\;\;\gamma=0.577\,.
\label{eq:13}
\eeq
For nuclei of finite radius $R_{A}$, Eq.~(\ref{eq:13}) holds
for $\rho\gsim R_{A}$, and $C(\rho\lsim R_{A})=C(R_{A})$.

Analytical evaluation of the path
integral (\ref{eq:7}) yields
\cite{BGZ}
\bea
S(\r_{2},\r_{2}',\xi_{2}|\r_{1},\r_{1}',\xi_{1})=
\left(\frac{\mu_{e}}{2\pi\Delta\xi}\right)^{2}
\exp\left\{\frac{i\mu_{e}}{2\Delta\xi}
[(\r_{1}-\r_{2})^{2}-(\r_{1}'-\r_{2}')^{2}]
\right.\nonumber\\
\left.
-\frac{1}{2}\int
d\xi n(\xi)
\sigma(|\ta_{s}(\xi)|)\right\}\,,
\label{eq:14}
\eea
$$
\ta_{s}(\xi)=(\r_{1}-\r_{1}')\frac{(\xi_{2}-\xi)}{\Delta\xi}
+(\r_{2}-\r_{2}')\frac{(\xi-\xi_{1})}{\Delta\xi}\,,\;\;\;
\Delta\xi=\xi_{2}-\xi_{1}\,.
$$
Introducing in (\ref{eq:8}) the Jacobi variables
$\bfa=(\mu_{e'}\r_{e'}+\mu_{\gamma}\r_{\gamma})/(\mu_{e'}+\mu_{\gamma})$
and $\bfb=\r_{e'}-\r_{\gamma}\,$, and integrating
over the trajectories $\bfa(\xi)$ and $\r_{e}(\xi)$ one can obtain
\bea
M(\r_{2},\r_{2}',\xi_{2}|\r_{1},\r_{1}',\xi_{1})=
\left(\frac{\mu_{e}}{2\pi\Delta\xi}\right)^{2}
\exp\left\{\frac{i\mu_{e}}{2\Delta\xi}
\left[(\r_{1}-\r_{2})^{2}-(\r_{1}'-\r_{2}')^{2}\right]
-\frac{i\Delta\xi}{l_{f}}\right\}\nonumber\\
\times
\int{\cal D}\bfb\exp\left\{i\int d\xi\left[
\frac{\mu_{e'\gamma}\dot{\bfb}^{2}}{2}+i\frac{ n(\xi)
\sigma(|\ta_{m}(\xi)|)}{2}\right]\right\}
\,,
\label{eq:15}
\eea
$$
\ta_{m}(\xi)=(\r_{1}-\r_{1}')\frac{(\xi_{2}-\xi)}{\Delta\xi}
+(\r_{2}-\r_{2}')\frac{(\xi-\xi_{1})}{\Delta\xi}
+\frac{\bfb(\xi)\mu_{\gamma}}{(\mu_{e'}+\mu_{\gamma})}\,,
$$
where $\mu_{e'\gamma}=\mu_{e'}\mu_{\gamma}/(\mu_{e'}+\mu_{\gamma})
=E_{e}x(1-x)$ ($E_{e}$ is the electron energy).

Substituting (\ref{eq:14}), (\ref{eq:15}) into (\ref{eq:6}),
and integrating in (\ref{eq:6})
over the transverse variables we finally obtain
(we set $-z_{1}=z_{2}=\infty$, and recover the
vacuum
term)
\beq
\frac{d P_{\gamma}}{d x}=2\mbox{Re}
\int\limits_{-\infty}^{\infty} d \xi_{1}
\int\limits_{\xi_{1}}^{\infty}d \xi_{2}
\exp\left(-\frac{i\Delta\xi}{l_{f}}\right)
g(\xi_{1},\xi_{2},x)\left[K_{e'\gamma}(0,\xi_{2}|0,\xi_{1})
-\left.K_{e'\gamma}(0,\xi_{2}|0,\xi_{1})\right|_{n=0}
\right]\,,
\label{eq:16}
\eeq
\beq
K_{e'\gamma}(\bfb_{2},\xi_{2}|\bfb_{1},\xi_{1})=
\int{\cal D}\bfb\exp\left\{i\int d\xi\left[
\frac{\mu_{e'\gamma}\dot{\bfb}^{2}}{2}-v(\bfb,\xi)\right]
\right\}\,\,,
\label{eq:17}
\eeq
\beq
v(\bfb,\xi)=-i\frac{n(\xi)\sigma(|\bfb |x)}{2}\,.
\label{eq:18}
\eeq
Thus, we expressed the intensity of the photon radiation
through the Green's function of the Schr\"odinger equation
with the imaginary potential (\ref{eq:18}).
Equation (\ref{eq:16}) is the main result of the present paper.

To proceed with analytical evaluation of the radiation
density we take advantage of the slow $\beta$-dependence
of $C(|\bfb|x)$ at $|\bfb|x\lsim 1/m_{e}$ which as we will show
below are important in Eq.~(\ref{eq:16}).
Evidently, to logarithmic accuracy we can replace
(\ref{eq:18}) by the harmonic oscillator potential
with the frequency
\beq
\Omega=
\frac{(1-i)}{\sqrt{2}}
\left(\frac{n C(\beta_{eff}x)x^{2}}{\mu_{e'\gamma}}\right)^{1/2}
=\frac{(1-i)}{\sqrt{2}}
\left(\frac{n C(\beta_{eff}x)x}{E_{e}(1-x)}\right)^{1/2}
\,\,.
\label{eq:19}
\eeq
Here $\beta_{eff}$ is the typical value of $|\bfb|$ for
trajectories contributing to the radiation density.
Making use of the oscillator Green's function, after some
algebra one can obtain from Eq.~(\ref{eq:16}) the
intensity of bremsstrahlung per unit length in the infinite medium
\beq
\frac{d P_{\gamma}}{dx dL}=
n\left(\frac{C(\beta_{eff}x)}{C(1/m_{e})}\right)
\left[
\left(\frac{d\sigma}{dx}\right)^{BH}_{nf}
S_{nf}(\eta)
+\left(\frac{d\sigma}{dx}\right)^{BH}_{sf}S_{sf}(\eta)
\right]\,.
\label{eq:20}
\eeq
Here $\eta=l_{f}|\Omega|$, and $(d\sigma/dx)^{BH}_{nf,sf}$ are
the Bethe-Heitler
cross sections conserving and changing the electron helicity.
The factors $S_{nf}$, $S_{sf}$ are given by
\beq
S_{nf}(\eta)=\frac{3}{\eta\sqrt{2}}
\int\limits_{0}^{\infty}
dy
\left(\frac{1}{y^{2}}-
\frac{1}{{\rm sh}^{2} y}\right)
\exp\left(-\frac{y}{\eta\sqrt{2}}\right)
\left[\cos\left(\frac{y}{\eta\sqrt{2}}\right)+
\sin\left(\frac{y}{\eta\sqrt{2}}\right)\right]\,.
\label{eq:21}
\eeq
\beq
S_{sf}(\eta)=\frac{6}{\eta^{2}}
\int\limits_{0}^{\infty}dy
\left(\frac{1}{y}-
\frac{1}{{\rm sh} y}\right)
\exp\left(-\frac{y}{\eta\sqrt{2}}\right)
\sin\left(\frac{y}{\eta\sqrt{2}}\right)\,.
\label{eq:22}
\eeq
At small $\eta$ $S_{nf}\simeq 1-16\eta^{4}/21$, and
$S_{sf}\simeq 1-31\eta^{4}/21$.
Up to the slowly dependent on $\eta$
factor $C(\beta_{eff}x)/C(1/m_{e})$, the suppression of bremsstrahlung
at $\eta\gg 1$ is controlled by the asymptotic behavior of the
factors (\ref{eq:21}), (\ref{eq:22}): $S_{nf}\simeq 3/\eta\sqrt{2}$,
$S_{sf}\simeq 3\pi/2\eta^{2}$.
Eqs.~(\ref{eq:21}), (\ref{eq:22}) allow to estimate $\beta_{eff}$.
The variable of integration in (\ref{eq:21}), (\ref{eq:22})
in terms of $\Delta \xi$ in Eq.~(\ref{eq:16}) equals $|\Delta \xi \Omega|$.
Therefore, for typical values of $\Delta \xi$ contributing to
the integral (\ref{eq:16}), $\Delta\xi_{eff}$,  we have
$
\Delta\xi_{eff}\sim {\rm min}(l_{f},1/|\Omega|)
$.
Then, $\beta_{eff}$ can be estimated from the
obvious Schr\"odinger diffusion relation
$\beta_{eff}\sim (2\Delta\xi_{eff}/\mu_{e'\gamma})^{1/2}$.
In the limit of low density, when $\eta \rightarrow 0$, we get
from this relation $\beta_{eff}\sim 1/m_{e}x$, and the right-hand
side of (\ref{eq:20}) reduces to the Bethe-Heitler cross section
times the target density.
In the soft photon
limit ($x\rightarrow 0$) at $n$ fixed,
Eqs.~(\ref{eq:20})-(\ref{eq:22}) yield
\beq
\frac{d P_{\gamma}}{dx dL}=
2\alpha^{2}Z\,\sqrt{\frac{2 n\log(2a/\beta_{eff}x)}{\pi E_{e}x}}\,,
\label{eq:24}
\eeq
where
$
\beta_{eff}\sim \left[
\pi (Z\alpha)^{2} n E_{e}x^{3}\log(2/\alpha Z^{1/3})\right]^{-1/4}\,.
$
Notice, that (\ref{eq:24}) differs from the prediction of
Ref. \cite{Migdal} which corresponds to
a replacement of  $\beta_{eff}x$ by $R_{A}$
in (\ref{eq:24}).

Let us now briefly consider a gluon bremsstrahlung by a quark
interacting with the color screened Coulomb centres.
Following Refs. \cite{GW,BD} we introduce a gluon mass $m_{g}=1/a_{QCD}$,
and treat the interaction of a quark with each centre in the Born
approximation.
The derivation of the gluon radiation rate follows closely the QED case.
Making use of the fact that $-T^{*}_{q}=T_{\bar{q}}$ (here $T_{q,\bar{q}}$
are the color generators for quark and antiquark)
and of the closure over final states of color centres, one can easily
show that in QCD one should only replace in (\ref{eq:18})
the QED dipole cross section $\sigma(|\bfb|x)$
by the QCD three-body cross section of scattering of the
color singlet $q\bar{q}g$
system on the color centre, $\sigma_{3}$ \cite{S3}. The latter in terms of the
color dipole cross section for scattering of a color singlet $q\bar{q}$ pair,
$\sigma_{2}$, is given by
\beq
\sigma_{3}(|\bfb|)=\frac{9}{8}
\left[
\sigma_{2}(|\bfb|)+\sigma_{2}(|\bfb|(1-x))
\right]-\frac{1}{8}\sigma_{2}(|\bfb|x)\,.
\eeq
For the formation length in QCD we have
$l_{f}=2E_{q}x(1-x)/[m_{q}^{2}x^{2}+m_{g}^{2}(1-x)]$.
The oscillator approximation is applicable for evaluation
of the  radiation density at $\eta_{QCD}\gg 1$.
In this regime $\beta_{eff}\ll a_{QCD}$, and to logarithmic
accuracy the intensity of gluon bremsstrahlung can be evaluated
making use of the oscillator Green's function with the
oscillator frequency
\beq
\Omega_{QCD}=
\frac{(1-i)}{\sqrt{2}}
\left(\frac{n C_{3}(\beta_{eff},x)}{\mu_{q'g}}\right)^{1/2}
=\frac{(1-i)}{\sqrt{2}}
\left(\frac{n C_{3}(\beta_{eff},x)}{E_{q}x(1-x)}\right)^{1/2}
\,\,.
\label{eq:25}
\eeq
Here
$$
C_{3}(\beta_{eff},x)=\frac{9}{8}[C_{2}(\beta_{eff})+
C_{2}(\beta_{eff}(1-x))]-\frac{1}{8}C_{2}(\beta_{eff}x)\,,
$$
and
$C_{2}(\rho)\simeq ({\alpha_{S}^{2}\pi C_{T}}/{6})
\log\left({2a_{QCD}}/{\rho}\right)$ ($C_{T}$ is the second
order Casimir for the color centre).

In QCD the limit $x\rightarrow 0$ at the same time corresponds
to $\eta_{QCD} \rightarrow 0$. Thus at sufficiently small $x$
the Bethe-Heitler regime takes place.
On the other hand, at $\eta_{QCD}\gg 1$ and $x\ll 1$ the oscillator
approximation yields
\beq
\frac{d P_{g}}{dx dL}=
2\alpha_{S}^{2}\sqrt{\frac{ nC_{T}\log(2a_{QCD}/\beta_{eff})}
{3\pi E_{q}x^{3}}}\,,
\label{eq:26}
\eeq
with
$
\beta_{eff}\sim [
x \alpha_{S}^{2}C_{T} n E_{q}]^{-1/4}\,.
$
Eq. (\ref{eq:26}) disagrees with
$dP_{g}/dxdL\propto x^{-3}$ predicted in Ref.~\cite{BD}.

I would like to thank N.N.~Nikolaev for stimulating
discussions and reading the manuscript of the paper.
I am grateful to J.~Speth for the hospitality at KFA,
J\"ulich, where this work was completed.

\end{document}